\documentclass[aps,pra,twocolumn,groupedaddress,showpacs]{revtex4-1}
\usepackage{graphicx}
\usepackage{dcolumn}
\usepackage{bm}
\usepackage{amsmath,amsfonts,amssymb,amsthm,verbatim,hyperref}
\usepackage[ansinew]{inputenc}
\usepackage[usenames]{pstcol}
\usepackage[normalem]{ulem}

\usepackage{color}
\usepackage{epsfig}

\newcommand{\be}{\begin{equation}}
\newcommand{\ee}{\end{equation}}
\newcommand{\bea}{\begin{eqnarray}}
\newcommand{\eea}{\end{eqnarray}}
\newcommand{\beq}{\begin{eqnarray}}
\newcommand{\eeq}{\end{eqnarray}}

\newcommand{\bi}{\begin{itemize}}
\newcommand{\ei}{\end{itemize}}
\newcommand{\bc}{\begin{center}}
\newcommand{\ec}{\end{center}}

\newcommand{\up}{\uparrow}
\newcommand{\down}{\downarrow}

\newcommand{\ra}{\rangle}

\def\be{\begin{equation}}
\def\ee{\end{equation}}
\def\bea{\begin{eqnarray}}
\def\eea{\end{eqnarray}}
\def\ba{\begin{array}}
\def\ea{\end{array}}
\def\pa{\partial}

\def\om{\omega}

\def\nn{\nonumber}
\def\ket{\rangle}
\def\bra{\langle}
\def\b{\mathbf}
\def\f{\frac}

\def\ra{\rightarrow}
\def\rm{\mathrm}

\def\bs{\boldsymbol}
\newcommand{\mv}[1]{\langle #1\rangle}
\newcommand{\dr}[1]{|#1\rangle}
\def\ua{\uparrow}
\def\da{\downarrow}

\begin{document}

\title{Quantum simulation of correlated-hopping models with fermions in optical lattices}
\author{M. Di Liberto$^{1}$, C. E. Creffield$^2$, G. I. Japaridze$^{3,4}$ C. Morais Smith$^1$}

\affiliation{$\mbox{}^1$ Institute for Theoretical Physics, Utrecht
University, Leuvenlaan 4, 3584CE Utrecht, the Netherlands}
\affiliation{$\mbox{}^2$Departamento de F\'isica de Materiales,
Universidad Complutense de Madrid, E-28040, Madrid, Spain}
\affiliation{$\mbox{}^3$ Andronikashvili Institute of Physics,
Tamarashvili 6, 0177 Tbilisi, Georgia} \affiliation{$\mbox{}^4$ Ilia
State University, Cholokasvili Avenue 3-5, 0162 Tbilisi, Georgia}

\begin{abstract}
By using a modulated magnetic field in a Feshbach resonance 
for ultracold fermionic atoms in optical lattices, we show that it is 
possible to engineer a class of models usually referred to as correlated-hopping 
models. These models differ from the Hubbard model in exhibiting 
additional density-dependent interaction terms that affect the hopping processes. 
 In addition to the spin-SU(2) symmetry, they also possess a charge-SU(2) symmetry, 
which opens the possibility of investigating the $\eta$-pairing mechanism 
for superconductivity introduced by Yang for the Hubbard model. We 
discuss the known solution of the model in 1D (where $\eta$ 
states have been found in the degenerate manifold of the ground state) and show
that, away from the integrable point, quantum Monte Carlo simulations at half filling predict 
the emergence of a phase with coexisting incommensurate spin and charge order.
\end{abstract}


\maketitle

\section{Introduction.}

The use of ultracold atoms in optical lattices as condensed
matter simulators has brought a major advance in physics in the last
decade. Both bosonic \cite{jaksch,greiner} and fermionic
\cite{hofstetter,esslinger} Hubbard models have been theoretically
and experimentally investigated, and the simulation of artificial
gauge fields \cite{dalibard,lihking} and Quantum Hall physics
\cite{aidelsburg,miyake,wouter} are some of the many phenomena that
this active field is unveiling \cite{lewenstein,bloch1,bloch2}.

The realization of the fermionic Hubbard model opens the possibility
of using quantum simulators to treat strongly-correlated fermionic
systems, with the ultimate goal of understanding high-$T_c$ 
superconductivity. While it is more challenging to cool fermionic
systems than bosonic ones, state-of-the-art techniques have recently
allowed fermionic atoms to be cooled sufficiently to reach the
regime where quantum magnetism manifests \cite{esslinger2}.

A particular interest with ultracold gases is the use of
time-dependent driving potentials. Using this technique, it has been
possible to observe the transition from a Mott-insulator to a
superfluid phase in the Bose-Hubbard model by a dynamical
suppression of tunneling \cite{eckardt2005,creffield,lignier2007},
as well as the simulation of frustrated classical magnetism
\cite{eckardt2010,struck2011}, and schemes for the realization of	
abelian \cite{struck2012} and non-abelian gauge fields
\cite{hauke2012}. More recently, a time-dependent modulation of
Feshbach resonances has been proposed for a system of ultracold
bosons, leading to a model with density-dependent hopping
coefficients, and exotic phenomena like pair superfluidity, and
holon and doublon condensation \cite{rapp2012}.

In this paper we extend this idea to fermionic atom systems. We
show how a time-dependent manipulation of the interaction strength
allows us to simulate an unusual class of ``correlated-hopping
models'' \cite{FF}, opening a window for the experimental
observation of a novel and elusive form of superconductivity
called $\eta$-superconductivity proposed by Yang in 1989. 
After discussing the model derivation and its symmetries,
we focus on the 1D case at half-filling and perform quantum 
Monte Carlo (QMC) simulations for arbitrary values of the 
Hubbard interaction $U$ and of the correlated-hopping parameter $\gamma$.
Our results show that the model can exhibit an interesting phase, with coexisting incommensurate spin- 
and charge-density-wave order.

\section{\label{model} Model}

We consider a system of (pseudo) spin-1/2 fermions in the lowest band
of an optical lattice, and use a Feshbach resonance to modulate the
interactions in time \cite{rapp2012}. The Hamiltonian of the model
reads
\be H = -J \sum_{\bra i,j\ket, \sigma} (c^\dag_{i\sigma}
c^{\phantom{\dagger}}_{j\sigma} + \rm{h.c.}) + \bar{U}(t) \sum_i
n_{i \ua} n_{i\da} \ee
where $J$ is the fermion hopping amplitude between
nearest-neighbor sites $\langle i, j \rangle$, and $\bar{U}(t)\equiv
U + U_1 \cos(\om t)$ is the time ($t$) dependent amplitude of the
two fermion coupling at the same site.

According to Floquet theory \cite{grifoni}, a time-periodic
Hamiltonian $H(t)=H(t+T)$ is described by a set of Floquet modes
$|u_n(t)\ket$ which are time-periodic with the same period $T$, and
a set of quasienergies $E_n$ which are solutions of the eigenvalue
equation
$$
\mathfrak{H}(t) \dr{u_{n}(t)}=E_n \dr{u_{n}(t)}\, ,
$$
where $\mathfrak{H}(t)\equiv H(t) - i \hbar \partial_t$ is called
the Floquet Hamiltonian. Solutions $\dr{\psi_n(t)}$ of the
Schr\"odinger equation thus have the form $\dr{\psi_n(t)} =
\exp(-iE_{n}t/\hbar)\dr{u_n(t)}$, and are unique up to a shift
$E^{\prime}_n= E_n + m\hbar\om$ of the quasienergies by an integer
multiple $m$ of $\hbar\om$, which thus gives a Brillouin-zone
structure in quasienergy. The eigenvalue problem is defined in the composite
Hilbert space \cite{sambe} $\mathcal{H'}=\mathcal{H} \otimes
\mathcal{H}_T $, where $\mathcal{H}$ is the standard Fock space and
$\mathcal{H}_T$ is the Hilbert space of time-periodic functions. Let
us define the following Floquet basis
\be | \left\{ n_{j\sigma} \right\}, m \ket = | \left\{ n_{j\sigma}
\right\} \ket e^{-i \f{U_1}{\hbar \om}\sin(\om t) \sum_j n_{j \ua}
n_{j\da}  + im\om t}\,, \ee
where $m$ labels the basis of the periodic functions, and $ | \left\{ n_{j\sigma}
\right\} \ket$ indexes the Fock states. The unitary transformation
performed by the operator $\exp [{-i (U_1/\hbar \om)\sin(\om t)
\sum_j \hat n_{j \ua} \hat n_{j\da}}]$, leads to the
time-independent Floquet Hamiltonian. The main goal is now the
calculation of the Floquet quasienergy spectrum, for which one needs
the matrix elements $\langle \mspace{-3mu}\langle \left\{
n_{j\sigma} \right\},m | \mathfrak{H}(t) | \left\{ n'_{j\sigma}
\right\},m' \rangle\mspace{-3mu} \rangle_{T}$. The symbol  $\langle
\mspace{-3mu}\langle \dots \rangle\mspace{-3mu} \rangle_{T}$ means
that the ordinary scalar product defined in $\mathcal H$ has been
time-averaged, defining the natural scalar product in $\mathcal H'$.
In the high-frequency regime, $\hbar\om \gg J,\,U$, states with
different label $m$ decouple, and the Floquet Hamiltonian matrix
elements can be approximated by $\langle \mspace{-3mu}\langle
\left\{ n_{j\sigma} \right\},m | \mathfrak{H}(t) | \left\{
n'_{j\sigma} \right\},m' \rangle\mspace{-3mu} \rangle_{T} \approx
\delta_{m,m'} \left( \langle \left\{ n_{j\sigma} \right\} |
H_{\textrm{eff}} | \left\{ n'_{j\sigma} \right\} \rangle + m \hbar
\om \delta_{n,n'} \right)$, defining an effective static Hamiltonian
\bea H_\rm{eff} &=&  -J \sum_{\bra i,j\ket,\sigma} (c^\dag_{i\sigma}
c^{\phantom{\dagger}}_{j\sigma} + \rm{h.c.})
\mathcal{J}_0 \left[K (n_{i\bar\sigma} - n_{j\bar\sigma}) \right] \nn\\
&&+  U \sum_i n_{i \ua} n_{i\da} \,. \eea
The function $\mathcal{J}_0 \left[K (n_{i\bar\sigma} -
n_{j\bar\sigma}) \right]$ is a Bessel function of the first kind.
Its argument is the density operator difference between sites $i$
and $j$ relative to the spin $\bar\sigma$, where
$\bar\sigma\equiv\,\, \da$ ($\ua$) if $\sigma=\,\, \ua$ ($\da$), and
the parameter $K=U_1/\hbar\om$.

We now perform a Taylor expansion of the Bessel function to rewrite
the hopping term. Using the fact that the Bessel function is an even
function, we can write its Taylor series (without necessarily
specifying the coefficients of the expansion) as
\bea \label{bessel} &&\mathcal{J}_0\left[K (n_{i\sigma} - n_{j\sigma}) \right]
=\sum_{m=0}^{\infty} c_{2m} K^{2m} (n_{i\sigma} -
n_{j\sigma})^{2m}\nonumber\\
&&= 1 + \left[ \mathcal{J}_0(K) - 1 \right] (n_{i\sigma} +
n_{j\sigma} - 2 n_{i\sigma} n_{j\sigma}) \, . \eea
In deriving (\ref{bessel}) we noted that the first term in the expansion
with $m=0$ is just $1$, and have used the fermion identity
$(n_{i\sigma} - n_{j\sigma})^{2m}=n_{i\sigma} + n_{j\sigma}
- 2 n_{i\sigma} n_{j\sigma}$ for arbitrary $m>0$. This allows the
Hamiltonian to be rewritten as
\begin{widetext}
\be \label{Eff_Ham_1} H_\rm{eff} = -J \sum_{\bra i,j\ket,\sigma}
(c^\dag_{i\sigma} c^{\phantom{\dagger}}_{j\sigma} + \rm{h.c.})
\left\{ 1 - X (n_{i\bar\sigma} + n_{j\bar\sigma} - 2 n_{i\bar\sigma}
n_{j\bar\sigma}) \right\}+  U \sum_i n_{i \ua} n_{i\da}\equiv  H_J +
H_U\,, \ee
\end{widetext}
where we define $X =1-\mathcal{J}_0(K)$.

Eq.\,(\ref{Eff_Ham_1}) can be easily recognized as the
Hamiltonian of the Hubbard model with a correlated-hopping
interaction \cite{Hub,FF}. Similar interaction terms have appeared in a different context in cold atoms. 
If one considers a fermionic lattice system very close to the Feshbach resonance 
(which is not the regime studied here) in a static magnetic field,
the behavior of the system  cannot be described using the one-band Hubbard model 
because the on-site interaction energy exceeds the energy gap and higher bands play an important role. The physics in this regime can be described by an effective one-band model with density-dependent tunneling rates \cite{duan2008,duan2010}. 

Let us now discuss some limits of the Hamiltonian (\ref{Eff_Ham_1}). In the absence of the
driving ($U_{1}=K=0$), the Bessel function $\mathcal{J}_0(K)=1$, 
and so the effective Hamiltonian (\ref{Eff_Ham_1}) coincides with the
Hamiltonian of the standard Hubbard model. Tuning the driving
to $K=2.4048..$, where $\mathcal{J}_0(K)=0$ and
$X=1$, produces a Hamiltonian that coincides, in $d=1$, with an
exactly solvable limit of the correlated-hopping model
(\ref{Eff_Ham_1}), in which the strongly correlated dynamics of the electrons
ensures separate conservation of the doubly occupied sites, 
empty sites, and singly occupied sites \cite{aligia}.
Interest in models with this particular type of fermionic dynamics
was triggered by the concept of $\eta$-superconductivity proposed by
C. N.~Yang \cite{yang}. Motivated by the discovery of high-$T_c$
superconductivity, Yang proposed a class of eigenstates of the
Hubbard Hamiltonian which have the property of off-diagonal
long-range order, which in turn implies the Meissner effect and flux
quantization \cite{yang2,GLS,nieh}, i.e. superconductivity. These
eigenstates are constructed in terms of operators $\eta^\dag_{\bs\pi} \equiv
\sum_\b{r} e^{- i \bs\pi\b{\cdot r}} c^\dag_{\b r\ua} c^\dag_{\b
r\da}$ that create pairs of electrons of zero size with momentum
$\bs\pi$. Yang also proved, however, that these states cannot be ground states
of the Hubbard model with finite interaction;
$\eta$-superconductivity is only realized in the Hubbard model at
infinite on-site attraction in $d\geq 2$ \cite{singh}. Later, several generalizations
of the Hubbard model showing $\eta$-superconductivity in the ground
state (for a finite on-site interaction) were proposed
\cite{essler2,essler,deboer,deboer2,schad}.

The exactly solvable limit of the model (\ref{Eff_Ham_1}) ($X=1$ in $d=1$)
has been analyzed in detail by Arrachea and Aligia
\cite{aligia,aligiaflux}. Away from the exactly solvable limit, the model
has been mainly studied in the weak-coupling limit ($X \ll J$) using the
continuum limit bosonization treatment and finite-chain exact diagonalization studies \cite{japaridze, aligia3}.

The infrared behaviour of the system (\ref{Eff_Ham_1}), determined
by the unusual correlated dynamics of fermions, is also strongly
influenced by its high symmetry. The three generators of the
spin-$\mathfrak{su}(2)$ algebra
\begin{eqnarray}\label{spin1}
S^{+} &=& \sum_{i}c^{\dagger}_{i\uparrow}c^{\phantom{\dagger}}_{i
\downarrow}, \, \, \ \
S^{-} = \sum_{i}c^{\dagger}_{i\downarrow}c^{\phantom{\dagger}}_{i\uparrow},\nonumber\\
S^{z} &=& {1 \over 2}\sum_{i}(n_{i\uparrow} - n_{i\downarrow}) ,
\end{eqnarray}
commute with the Hamiltonian (\ref{Eff_Ham_1}), which shows its
spin-$SU(2)$ invariance.

To keep the discussion as general as possible, let us consider the case
of bipartite lattices that we label $A,\,B$ and introduce the index $\alpha_i$ that
assumes values $\alpha_i = 1$ if $i\in A$ and $\alpha_i=-1$ if $i\in B$. In $d=1$,
in particular, one can choose $A$ to be the even sites and $B$ to be the odd sites and
one simply has $\alpha_i = (-1)^i$.
The electron-hole transformation $c_{i, \sigma} \rightarrow
(-1)^{\alpha_i}c^{\dagger}_{i, \sigma}$ 
leaves the Hamiltonian unchanged
and therefore the model is characterized by the electron-hole
symmetry. Moreover, for the case of half-filling that
we consider in this work, the
model (\ref{Eff_Ham_1}) possesses an additional
spin-$SU(2)$ symmetry. The transformation
\begin{eqnarray}\label{spinEHtransf}
c_{i\uparrow} &\rightarrow& c_{i\uparrow}\, , \qquad c_{i,
\downarrow} \rightarrow  (-1)^{\alpha_i}\,c^{\dagger}_{i\downarrow},
\end{eqnarray}
interchanges the charge and spin degrees of freedom and  converts
\begin{equation}\label{ehhamilt2}
{H_\rm{eff}}(J, U , X) \rightarrow {H_\rm{eff}}(J, -U, X).
\end{equation}
In this case therefore, the charge sector is governed by the same
$SU(2)$ symmetry as the spin sector, and the model has the
$SU(2)\otimes SU(2)$ symmetry \cite{aligia} with generators
\begin{eqnarray}\label{spin2}
&\eta^{+} =
\sum_{i}(-1)^{i}c^{\dagger}_{i\uparrow}c^{\dagger}_{i\downarrow}, \
\, \ \
\eta^{-} = \sum_{i}(-1)^{i}c_{i\downarrow}c_{i\uparrow},&\nonumber\\
&\eta^{z} = {1 \over 2}\sum_{i}(1-n_{i,\uparrow}-n_{i,
\downarrow}).&
\end{eqnarray}

Henceforth we will focus on the case $d=1$ where an exact solution
of the model exists both for $X=0$ (Hubbard model) and $X=1$, as previously mentioned.
For the half-filled Hubbard model the $SU(2) \otimes SU(2)$ symmetry
implies that the gapped charge and the gapless spin sectors for
$U>0$ are mapped by the transformation Eq. (\ref{spinEHtransf}) into
a gapped spin and a gapless charge sector for $U<0$. Moreover, at
$U<0$ the model is characterized by the coexistence of CDW and
singlet superconducting (SS) instabilities in the ground state
\cite{FK}.

Contrary to the on-site Hubbard interaction $U$, the $X$ term
remains invariant with respect to the transformation Eq.\,(\ref{spinEHtransf}). 
For a given $X$, this immediately implies that, 
\begin{itemize}
\item
for $U=0$ the properties of the charge and the spin sectors are
identical;
\item
in the limit in which $U\gg X$ one expects that the large onsite repulsion would open a gap in the charge sector. Since for $U=0$ the spin and charge degrees of freedom have the same properties because of the $X$ symmetry, there must exist a critical value of the Hubbard coupling $U_{c}\geq 0$ corresponding to a crossover from the $X$ dominated regime
into a $U$ dominated regime.
\item
The Luttinger-liquid parameters of the model characterizing the gapless charge
($K_{\rho}$) and spin ($K_{\sigma}$) degrees of freedom are $K_{\rho}=K_{\sigma}=1$.
\end{itemize}

In the following sections, we will separately consider the exactly
solvable cases ($X=0$ and $X=1$), and the physically relevant case of
($0<X<1$).

\section{Exactly solvable case: $X=1$, $d=1$.}

In this section, we mainly follow the route developed by
Arrachea and Aligia \cite{aligia}. At $X=1$, the hopping of an electron
with spin $\sigma$ from a site $i$ to a neighboring site $j$ is only possible
if there are no other particles on the sites
($n_{i\bar{\sigma}}=n_{j\bar{\sigma}}=0$), or if both sites are occupied
by electrons with opposite spin
($n_{i\bar{\sigma}}=n_{j\bar{\sigma}}=1$). Thus, the only allowed
hopping processes in this limit are exchange processes
of a singly occupied site with a holon 
(e.g. $|0, \up \ket \leftrightarrow | \up, 0 \ket$),
and a doublon with a singly-occupied site
(e.g. $| \up \down, \up \ket \leftrightarrow | \up, \up \down \ket$). 

It is convenient to use  the slave-particle formalism to rewrite
the model in another basis, where all available processes are
clearly displayed. One defines the mapping
\be
|0\ket_j \ra h^\dag_j |0\ket \, ,\,\, |\sigma\ket_j \ra f^\dag_{j
\sigma} |0\ket \,, \,\, \left|\,\uparrow\downarrow\right.\ket_j \ra
d^\dag_j |0\ket \, ,
\ee
where the slave particles must obey the constraint
\be
h^\dag_i h_i + d^\dag_i d_i + \sum_\sigma f^\dag_{i\sigma}
f_{i\sigma} = 1
\ee 
at each lattice site. The constraint physically
means that the slave particles act as hard-core particles,
with an infinitely large on-site repulsion. The $h^\dag_i$ and
$d^\dag_i$ bosonic operators describe, respectively, holons and
doublons of the original system, while the fermionic operators
$f^\dag_{i\sigma}$ describe fermions with spin $\sigma$. Using this
mapping, the Hamiltonian can be exactly rewritten in the form
\bea \label{hamsl} H^0_\rm{eff} &=&-J \sum_{\mv{i,j} , \sigma}
[f^\dag_{j\sigma}f_{i,\sigma} (h^\dag_i h_j - d^\dag_i d_j) +
\rm{h.c.}] \nn \\&+& U\sum_i d^\dag_i d_i\,, \eea
where one can immediately observe that the numbers $N_h$, $N_d$,
$N_{f\ua}$ and $N_{f\da}$ are {\em separately} conserved, because the
Hamiltonian (\ref{hamsl}) can only interchange individual particles.
This corresponds to a $U(1)$ symmetry for each slave particle
sector. These particle numbers will therefore be used as quantum
numbers to label the eigenstates. Notice that the $U$ term plays the
role of a chemical potential for doublons and that there is not, in
general, a free part of the Hamiltonian for the slave particles. We
stress the sign difference in the exchange process between doublons
and holons in Eq.\,(\ref{hamsl}). The additional minus sign for the
doublons is responsible for the $\eta$-symmetry with momentum
$\bs\pi$. While the restricted dynamics of hopping processes expressed 
in the Hamiltonian (\ref{hamsl}) is a very general property of the choice 
$X=1$, it is only in $d=1$ that there are additional symmetries (not discussed here) 
that allow the model to be solved exactly. The solution of this model in 1D for open boundary
conditions was given in Ref.\,\cite{aligia}. 
The physical properties of the 1D system described by
Eq.\,(\ref{hamsl}) are very peculiar. When a doublon and a holon are
neighbors, they act like hard-core bosons as previously mentioned,
and cannot tunnel through each other because of the dimensionality
of the system. Such a process would require the doublon and the
holon to annihilate into two single fermions on the neighboring
sites and then reform as a doublon and holon on
exchanged sites. This process is forbidden at $X=1$, but is
possible for $X \neq 1$.

\begin{figure}[!tbp]
\begin{center}
\includegraphics[width=0.4\textwidth]{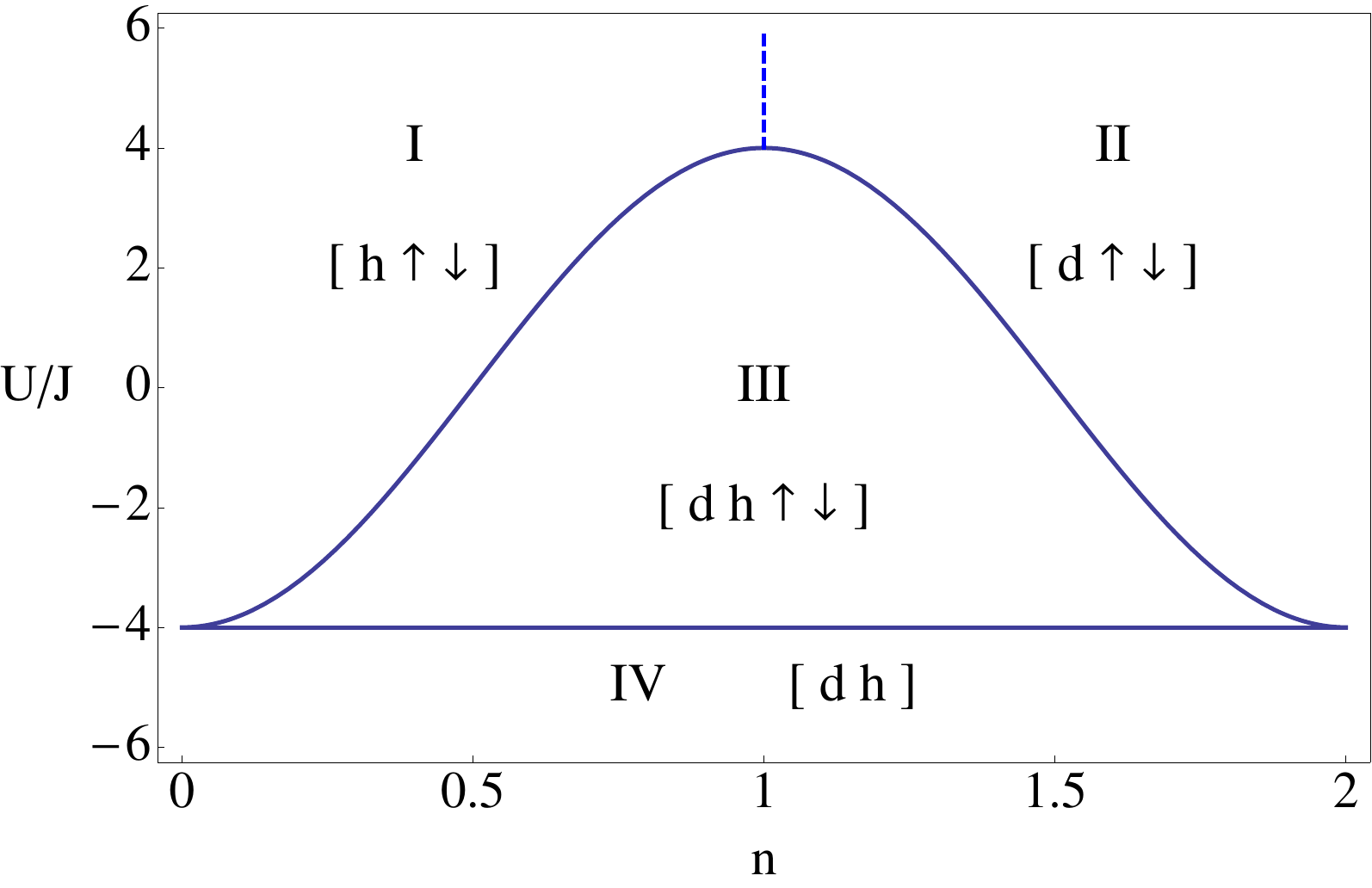}
\end{center}
\caption{(Color online) Phase diagram of model (\ref{Eff_Ham_1}) obtained for $\mathcal{J}_0(K)=0$ in $d=1$ \cite{aligia}. }
\label{phdia}
\end{figure}

As a result, there are three regimes for the ground state phase diagram, as shown in
Fig.\,\ref{phdia}: in region I there are only single fermions 
and holons (in region II, by particle-hole symmetry, only doublons and 
single fermions); the dashed line in Fig.\,\ref{phdia} will be discussed later when we concentrate
on the regime $n=1$; in region III all three types of particles are present, single
fermions, holons and doublons; in region IV there are no single fermions but only 
doublons and holons. In all sectors the ground state is highly degenerate
and, in region III and IV, one can show that also $\eta$-states belong to the
ground state manifold. For further details, we refer the reader to Ref.\,\cite{aligia}.

In the following, we focus on the half-filled case $n=1$, for which we perform  
QMC simulations in Sec.\,V.
Since at $X=1$ and half-filling the number of doublons $N_{d}$, holons $N_{h}=N_{d}$
and of the single occupied sites $N_{f}=N-2N_{d}$ are integrals
of motion, the delocalization energy of the system coincides with
that of $2N_{d}$ hard-core bosons on a lattice of $N$ sites. This
equivalence allows one to write the density of energy at half-filling as
\be \epsilon(n_d) = -\f{2J}{\pi} \sin\left(\pi -2\pi n_d\right) +
U\,n_d\,. \ee
For the half-filled case, the three regimes mentioned before become
(see Fig.\,\ref{phdia}):

\emph{i}.\, $U<-4J$. In this case, the ground state only contains
doublons and holons ($n_d=n_h=1/2$) that are frozen in the ground state since no dynamics is
allowed in the absence of single fermions. The system is a doublon-holon insulator 
and its energy is $E_{0}=NU/2$. The degeneracy of the ground state $\sim N$ diverges 
in the thermodynamic limit.

\emph{ii}.\, $U>4J$. In this case there are no doublons and holons in the
ground state; all sites are singly occupied and particles cannot
hop. The ground state has energy  $E_{0}=0$ and is
$2^{N}$ times degenerate, due to the freedom of distribution of spins
of particles along the lattice. This state is thus a charge insulator. 

\emph{iii}.\, $-4J \leq U \leq 4J$. In this case, the ground state consists
of a finite number of doublons and holons, separated by singly occupied
sites to ensure their maximal delocalization along the lattice. It is clear that at 
$U=0$ the minimum of kinetic energy is reached at the densities $n_{d}=n_h=0.25$.
The doublon density now depends on the ratio $U/J$:
\be n_d=\f{1}{2}\left[1-\f{1}{\pi}\arccos\left(-\f{U}{4J}\right) \right]\,. \ee


\section{Away from the exact solution.}

Deviation from the
exactly solvable limit $X=1$ produces the new term
\be \label{hamgamma}  2\gamma\, J \sum_{\bra i,j \ket, \sigma}
[f^\dag_{i\ua} f^\dag_{j\da} (e_i d_j + e_j d_i) + \rm{h.c.} ]\,,
\ee
where we have defined $\gamma\equiv\mathcal{J}_0(K) = 1 - X$. As these new
terms allow doublons to convert into two single fermions on
neighboring sites (and the reverse), the Hamiltonian no longer
conserves the individual number of slave particles, and thus no
exact solution is known. However, the $\eta$-symmetry {\em is}
preserved and one expects that the enormous degeneracy of the ground
state would again be removed. For $\gamma \simeq 1$, {\it i.e.} $|X|\ll
1$, the model can be treated using bosonization techniques and the
phase diagram is known (also in the presence of nearest-neighbor
interactions) \cite{japaridze} showing for $U<0$ that superconducting
correlations coexist with CDW. For the strongly-interacting case,
exact  diagonalization in 1D has been used \cite{arrachea,nakamura},
for systems of up to 12-14 sites. 
Nakamura \cite{nakamura} presented a phase diagram at
half and quarter filling (for $X=\pm 1/4$), and Arrachea \emph{et al.}
\cite{arrachea} have shown that superconducting correlations can
appear for $n=1$.
However, the general picture for arbitrary
filling is still missing.

In the next section we will use the QMC technique to investigate the 
charge and spin ordering
of the Hamiltonian given by Eq.\,(\ref{Eff_Ham_1}) at general
values of $X$ and see how
the results evolve between the two integrable
cases ($X=0$ and $X=1$) for the specific case of half-filling.

\section{Quantum Monte Carlo method}

To treat the Hamiltonian (\ref{Eff_Ham_1}), we employed a standard ``world-line''
algorithm \cite{worldline}. This is a finite-temperature method,
operating in the canonical ensemble, which is particularly
well-adapted to treat lattice spin-charge Hamiltonians. In order to sample the zero
temperature behavior of the system, it is important to set the inverse temperature of
the system, $\beta = 1/k T$, to a sufficiently large value. By comparing the results
for the ground state energy of the system with $\gamma=1$ to the exact
results for the Hubbard model available from the Bethe Ansatz, we established that
a sufficiently low temperature was $\beta J = 48$, and accordingly we used
this value in all the simulations. The Trotter decomposition of the imaginary
time axis gives systematic errors which can be made arbitrarily small by
increasing the number of timeslices, thereby reducing the imaginary-time
discretization $\Delta \tau$. Our simulations demonstrated that the convergence of
the results depended strongly on the value of $\gamma$. For the Hubbard model
($\gamma=1$) a relatively coarse value of $\Delta \tau = 0.1$ was adequate. However,
as $\gamma$ was reduced, $\Delta \tau$ also had to be reduced further, the lowest values
of $\gamma = 0.2$ requiring a discretization of $\Delta \tau \simeq 0.02$, with
the simulation involving 2048 timeslices.

As well as the increased number of timeslices required, taking lower values of 
$\gamma$ was also hindered by ergodic ``sticking'', in which local Monte Carlo (MC) updates are
unable to evolve the system from local minima in energy. It was this factor that
set the practical barrier on the lowest values of $\gamma$ that we were able to simulate,
and accordingly we only present results for $\gamma \geq 0.2$. In order to obtain
results of high accuracy, typically 16,000 MC measurements would be made for each
set of parameters, with each measurement being separated by the next by
several MC sweeps in order to reduce autocorrelation between the data.
 
A particular advantage of the world-line method is that as it operates in
the real-space occupation number basis, it is simple to
evaluate operators diagonal in number operators, such as the onsite
spin $\sigma_i = n_{i \ua} - n_{i \da}$, the onsite charge
$\rho_i = n_{i \ua} + n_{i \da}$, the doublon number, and correlations
between these operators. An especially useful quantity is the static
structure function
\be
S_{\alpha}(q) = \frac{1}{L} \sum_{m,n} e^{i q \left(m - n \right)}
\bra \alpha_m \alpha_n - \alpha^2 \ket
\label{structure}
\ee
where $m$ and $n$ are integers labeling sites, $\alpha$ = $\sigma$ ($\rho$) 
denotes spin (charge), and $L$ is the number of lattice sites. 
As well as using the structure functions to investigate the type
of spin and charge ordering present in the system, they can also be used to
directly estimate the Luttinger-liquid parameters \cite{clay},
\be
K_{\alpha} = \mbox{lim}_{q \rightarrow 0} \frac{S_{\alpha}(q)}{\pi q}\,.
\label{lutt_par}
\ee
Thus, when the structure function is linear at low momentum, the Luttinger-liquid parameter
is well-defined and is simply related to its slope. On the other hand, if the function is
quadratic, this indicates that the Luttinger-liquid parameter is not well-defined
and that this sector has a gap. 
In a uniform system, continuity requires that $S(q \rightarrow 0) = 0$. 
However, if phase separation
occurs $S(q)$ will have a peak at the smallest non-zero momentum, which
will diverge as $L$ increases. The regularity of $S(q)$ for small momentum
thus also provides a first check that the system is not phase-separated.

\subsection{\label{qmc_results} QMC results}

\begin{figure}
\begin{center}
\includegraphics[width=0.45\textwidth,clip=true]{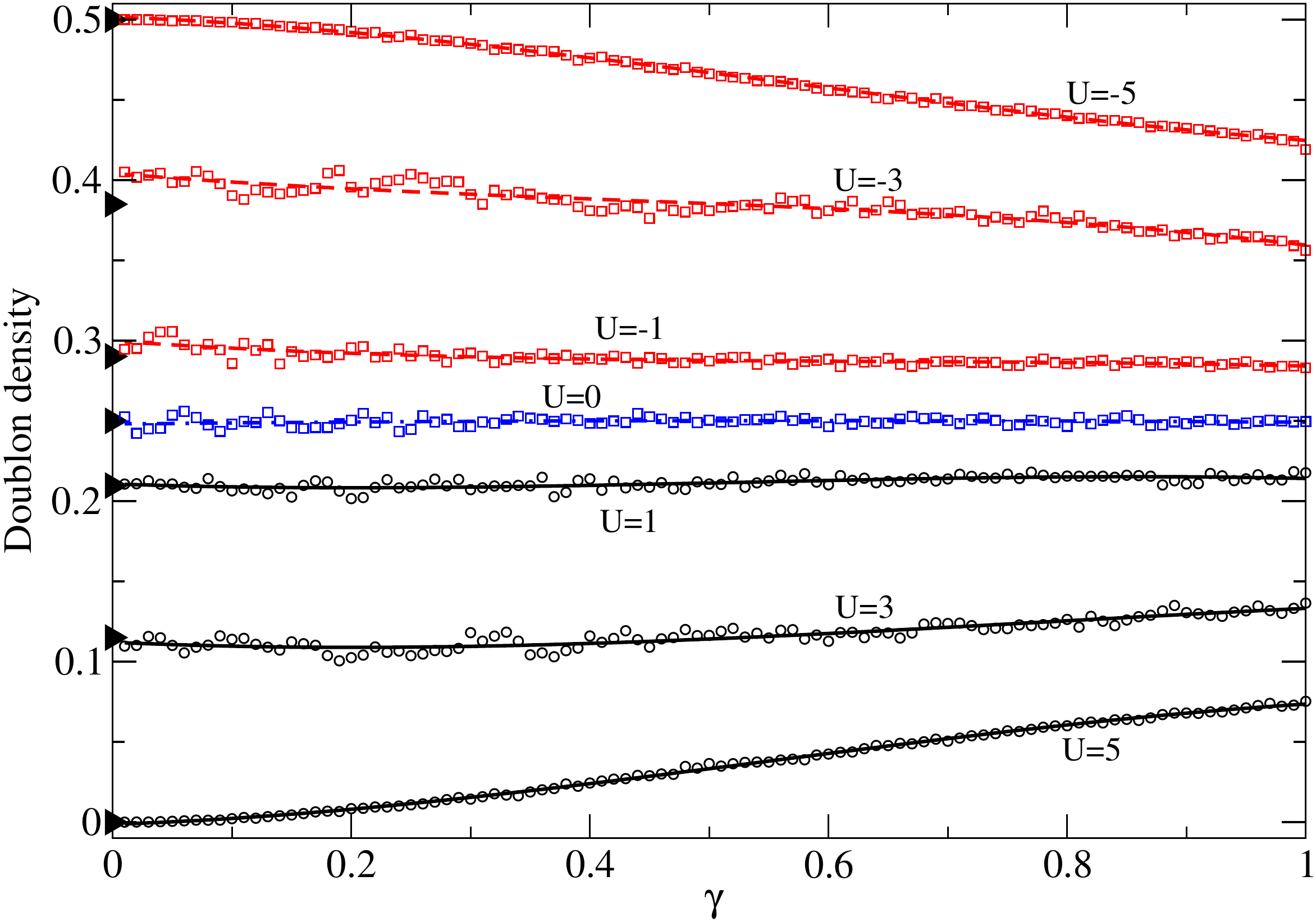}
\end{center}
\caption{(Color online) Doublon density measured as $\gamma$ is quasi-statically reduced
toward zero. For $U=0$ (blue squares) the doublon density does not change,
reflecting the symmetry of the model. For positive $U$ (black circles) the
density drops as $\gamma$ is reduced; the reverse is the case for negative $U$ (red squares).
The data are symmetric about $n_d = 0.25$ (see Eq.\,(\ref{reflect})). The solid lines are
cubic fits to the data to guide the eye. The arrows on the left indicate
the analytic values obtained in Ref.\,\cite{aligia} for the limit
$\gamma=0$.
Parameters of the model: 32 sites,
$\beta = 48$, $n_\ua = n_\da$.}
\label{doublon_density}
\end{figure}

{\em Doublon density -- }
In this section we will measure all energies in units of $J$. 
In Fig.\,\ref{doublon_density} we show the doublon density as a function of $\gamma$
for several different values of the Hubbard interaction $U$. In these simulations,
$\gamma$ was initially set to 1, and then reduced ``quasi-statically'' in steps of
$\Delta \gamma = 0.01$. For each value of $\gamma$ and $U$ the ensemble was
allowed to rethermalize and a number (typically 64) of MC measurements made. 
This technique permits a rapid scan to be made through the configuration space of the
model, at the expense of only producing results of moderate accuracy. 

From Fig.\,\ref{doublon_density}, we can first see that for $U=0$, the doublon
density does not depend on $\gamma$. This arises from the underlying symmetry of 
the Hamiltonian
at half-filling. For negative $U$, we see that the doublon density increases as
$\gamma$ is reduced, interpolating smoothly between the results for the
Hubbard model ($\gamma=1$), and the exactly solvable case ($\gamma=0$).
The results for positive $U$ mirror those for negative $U$, and can be related via
\be
n_d(+|U|) = 0.5 - n_d(-|U|).
\label{reflect}
\ee
The validity of Eq.\,(\ref{reflect}) is clear from the numerical results in Fig.\,\ref{doublon_density}. In addition, 
it can be easily proven, starting from Eq.\,(\ref{spinEHtransf})
and recalling that $N_d = \sum_i n_{i\ua} n_{i\da}$.

\begin{figure}
\begin{center}
\includegraphics[width=0.45\textwidth,clip=true]{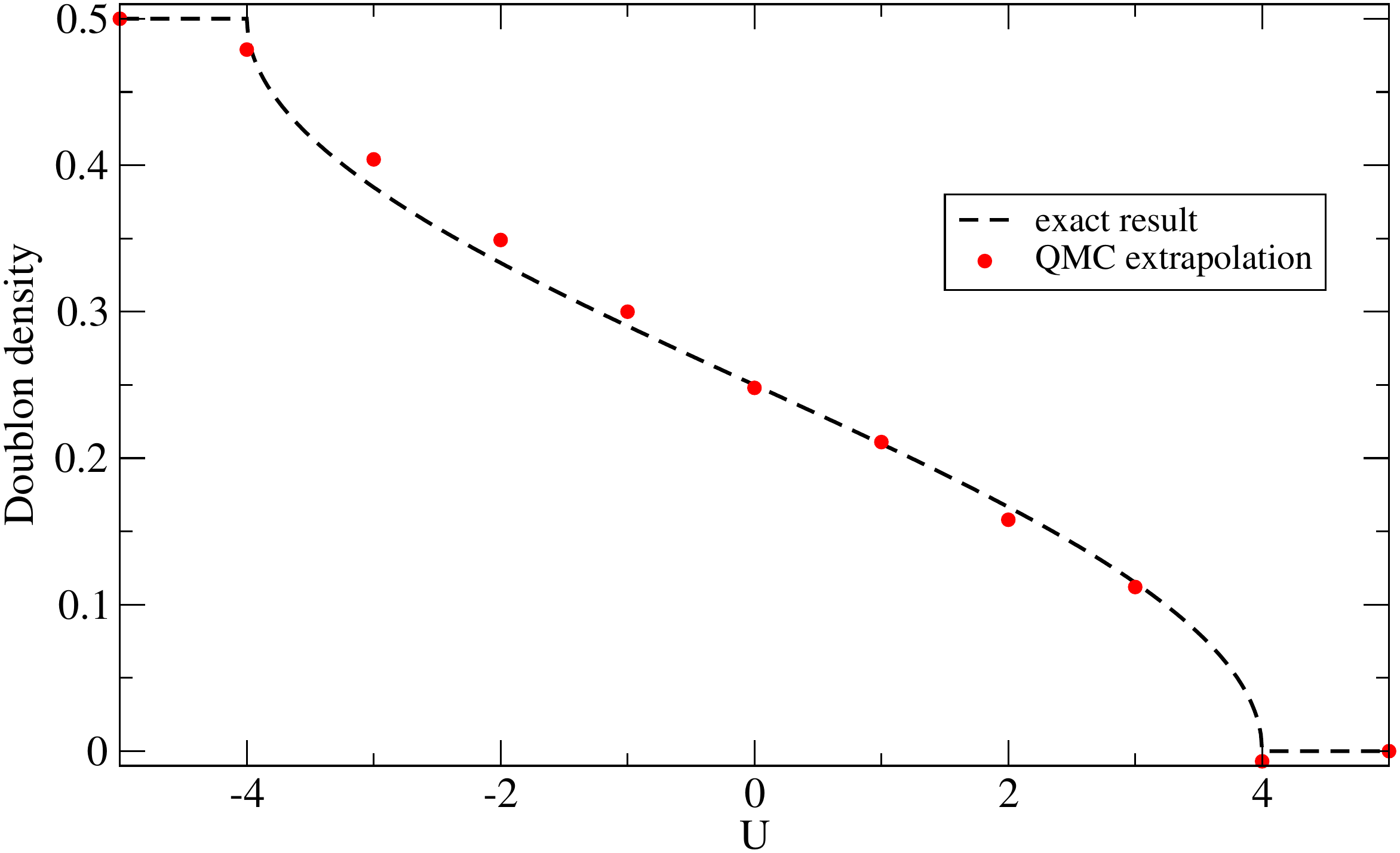}
\end{center}
\caption{(Color online)  Comparison of the analytic results for the doublon
density for $\gamma=0$ with the results obtained by extrapolating the data shown
in Fig.\,\ref{doublon_density}. The agreement is seen to be excellent.}
\label{compare_aligia}
\end{figure}

Although $\gamma = 0$ is not directly accessible to our QMC simulation due to ergodic trapping,
we can obtain estimates for the doublon density in this limit by extrapolating
the data in  Fig.\,\ref{doublon_density}. We present the results in Fig.\,\ref{compare_aligia}.
The agreement between the numerical results and the exact solution \cite{aligia} is 
excellent, demonstrating the accuracy and reliability of the QMC simulation.

\begin{figure}
\begin{center}
\includegraphics[width=0.45\textwidth,clip=true]{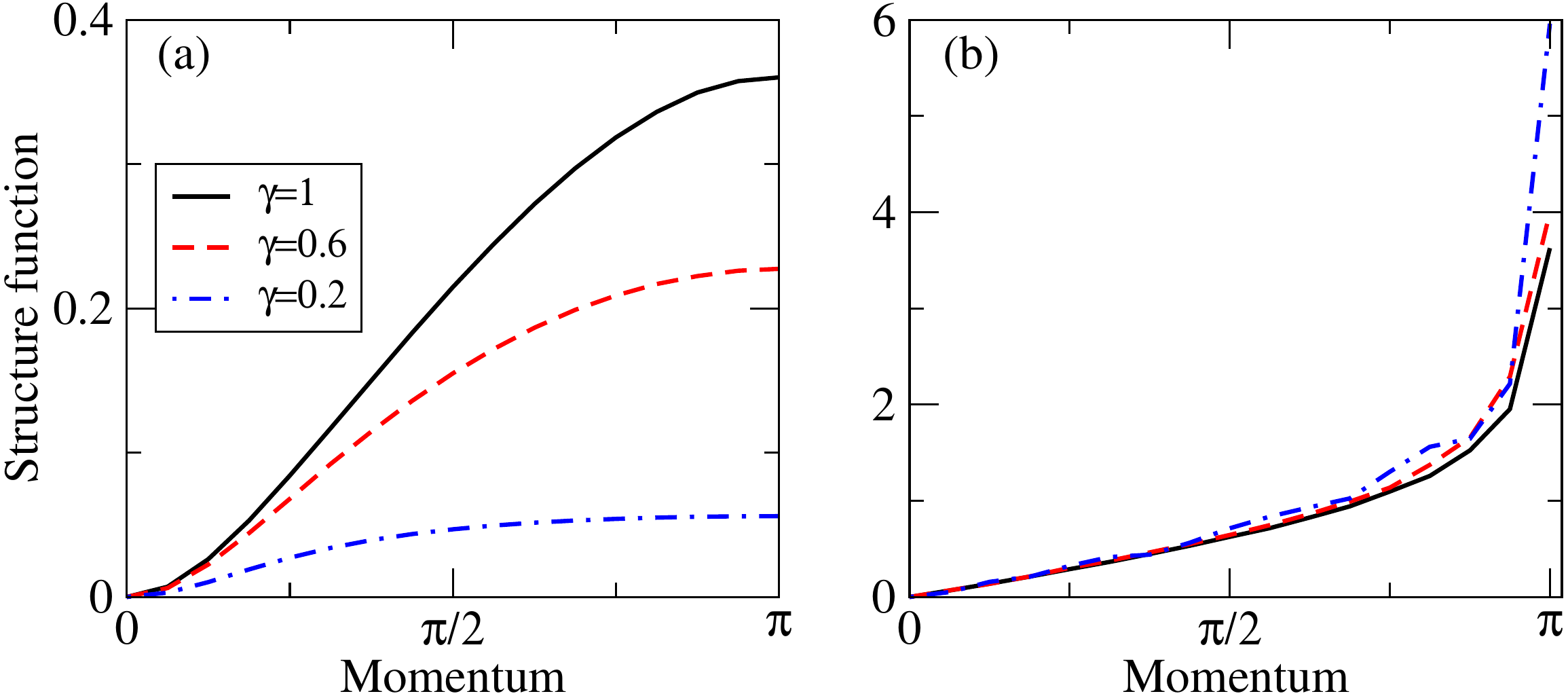}
\includegraphics[width=0.45\textwidth,clip=true]{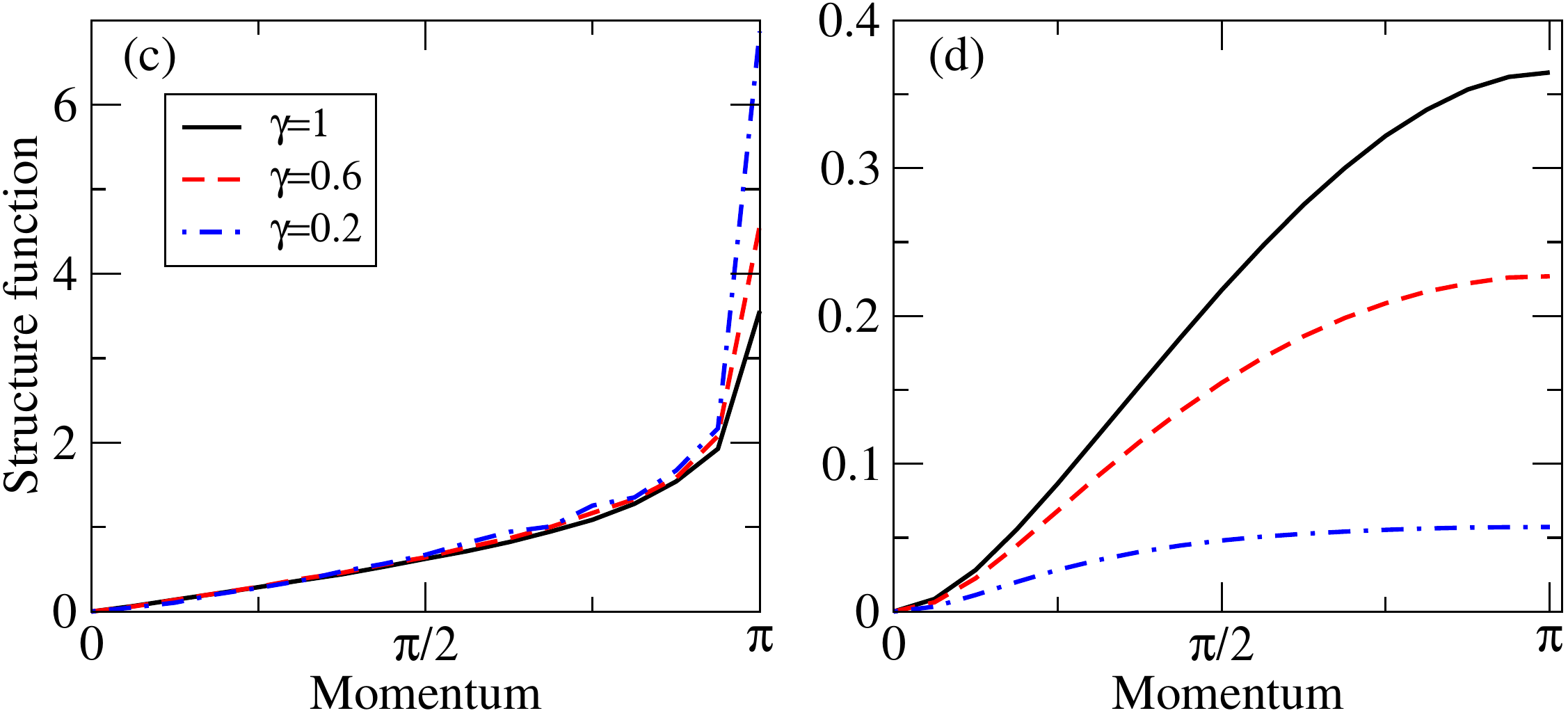}
\end{center}
\caption{(Color online) (a) Charge structure function for the half-filled model
with $U=4$. For $\gamma=1$ there is a weak peak at $k=\pi$; as $\gamma$
is reduced this peak is suppressed.
(b) Spin structure function. For $\gamma=1$ there is a strong peak at
$k=\pi$ indicating strong antiferromagnetic ordering. This peak
is enhanced as $\gamma$ is reduced.
(c) As above, but for $U=-4$. (d) As above, but for $U=-4$. The effect
of changing the sign of $U$ is to interchange the spin and charge degrees of freedom.
Momentum is measured in units of the inverse lattice spacing.}
\label{mc_U4}
\end{figure}

{\em Correlation functions -- }
In Fig.\,\ref{mc_U4}a we show the static charge structure functions
for strong repulsive interactions, $U=4$. It can be clearly seen that for the Hubbard model
($\gamma=1$) the charge sector is gapped, and that the structure function
presents a weak peak at $k = 2 k_F = \pi$. Reducing $\gamma$ suppresses
the structure function, and weakens this peak further. The spin
structure function, shown in Fig.\,\ref{mc_U4}b shows a contrasting behavior.
For the Hubbard model this function possesses a strong peak at $2 k_F$,
indicating the presence of strong antiferromagnetic ordering
$(\ua, \da, \ua, \da )$, and this peak
is enhanced as $\gamma$ is reduced. An infinitesimally
small deviation of the coupling $\gamma$ from zero
opens channels for the exchange of spins on neighboring sites.
This gives a preference for
an alternating distribution of particles with opposite spins along the
lattice, \emph{i.e.} a spin-density-wave (SDW) structure.
The spin excitations are gapless, and
the spin-$SU(2)$ symmetry sets the Luttinger-liquid parameter
$K_{\sigma}=1$.

Below these plots we show the corresponding structure functions for attractive
interaction $U=-4$. It can be clearly seen  that changing the sign of $U$ simply has
the effect of interchanging the spin and charge degrees of freedom,
as noted in Section \ref{model}. In this case we see that reducing $\gamma$ now has
the effect of suppressing the spin dynamics, while enhancing the staggered charge
order $(d, h, d, h)$. Similarly to before, deviation from the exact solution for $\gamma=0$
opens channels for 
an alternating distribution of doublons and holons along the
lattice i.e. a charge-density-wave (CDW) structure. Now the spin
degrees of freedom are gapped, the charge excitations are gapless,
and due to the charge-$SU(2)$ symmetry, the Luttinger-liquid
parameter $K_{\rho}=1$.

\begin{figure}
\begin{center}
\includegraphics[width=0.45\textwidth,clip=true]{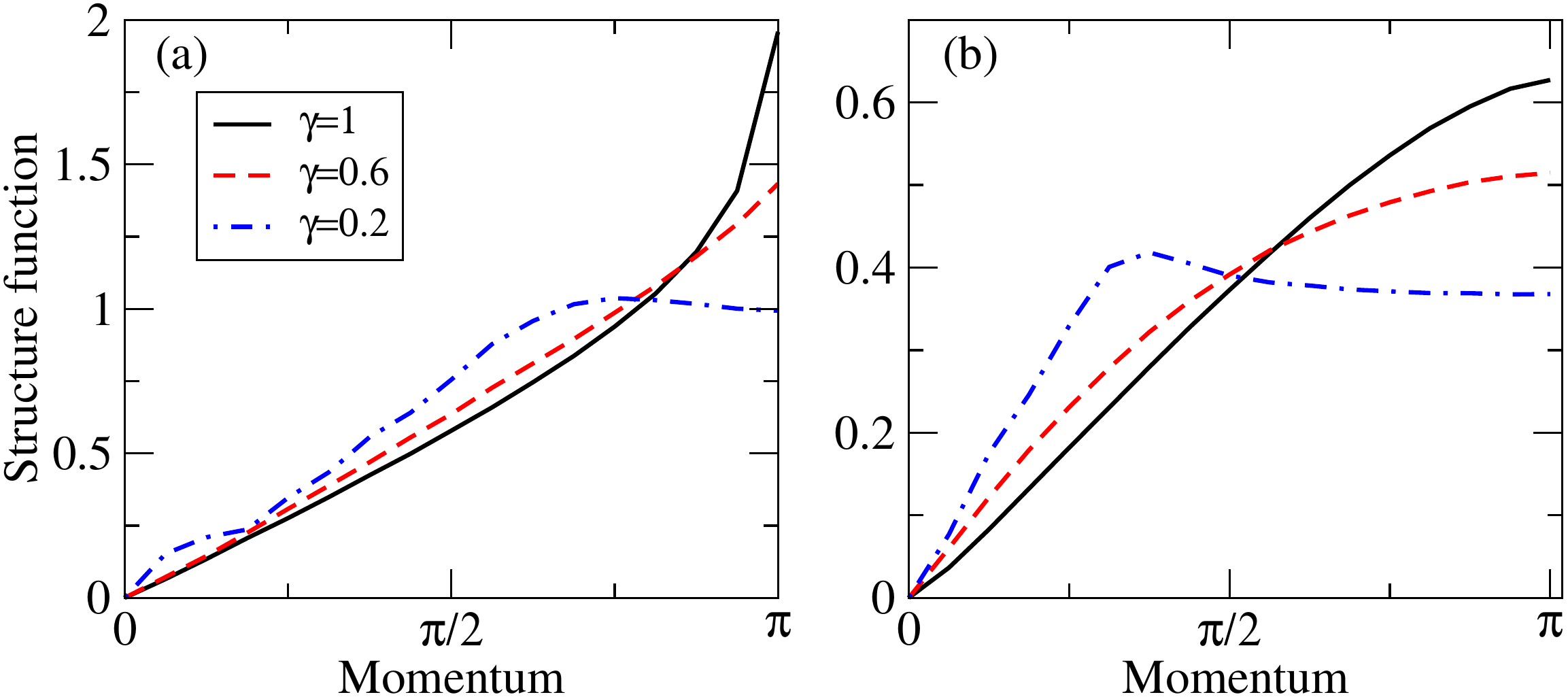}
\end{center}
\caption{(Color online) (a) Charge structure function for $U=-2$. As $\gamma$ is reduced
below 0.6, the system forms an incommensurate CDW.
(b) Spin structure function for $U=-2$. For low $\gamma$ the system also forms
an incommensurate SDW. Momentum is measured in units of the inverse lattice spacing.}
\label{mc_U-2}
\end{figure}

Results for $U=-2$ are given in Fig.\,\ref{mc_U-2}. Looking first at
the charge structure function, the result for the Hubbard model looks similar
to that seen previously for $U=-4$. As $\gamma$ is reduced, however, a new behavior emerges.
When $\gamma$ is reduced below 0.6, the charge structure function forms
a peak at an {\em incommensurate} momentum, indicating the formation of an
incommensurate CDW. At the same time an incommensurate SDW
forms in the spin sector, at a smaller value of momentum. 
This incommensurate ordering is reminiscent of the behavior known for
stripes in the 2D conventional Hubbard model upon doping
\cite{zaanen}. 

\begin{figure}
\begin{center}
\includegraphics[width=0.45\textwidth,clip=true]{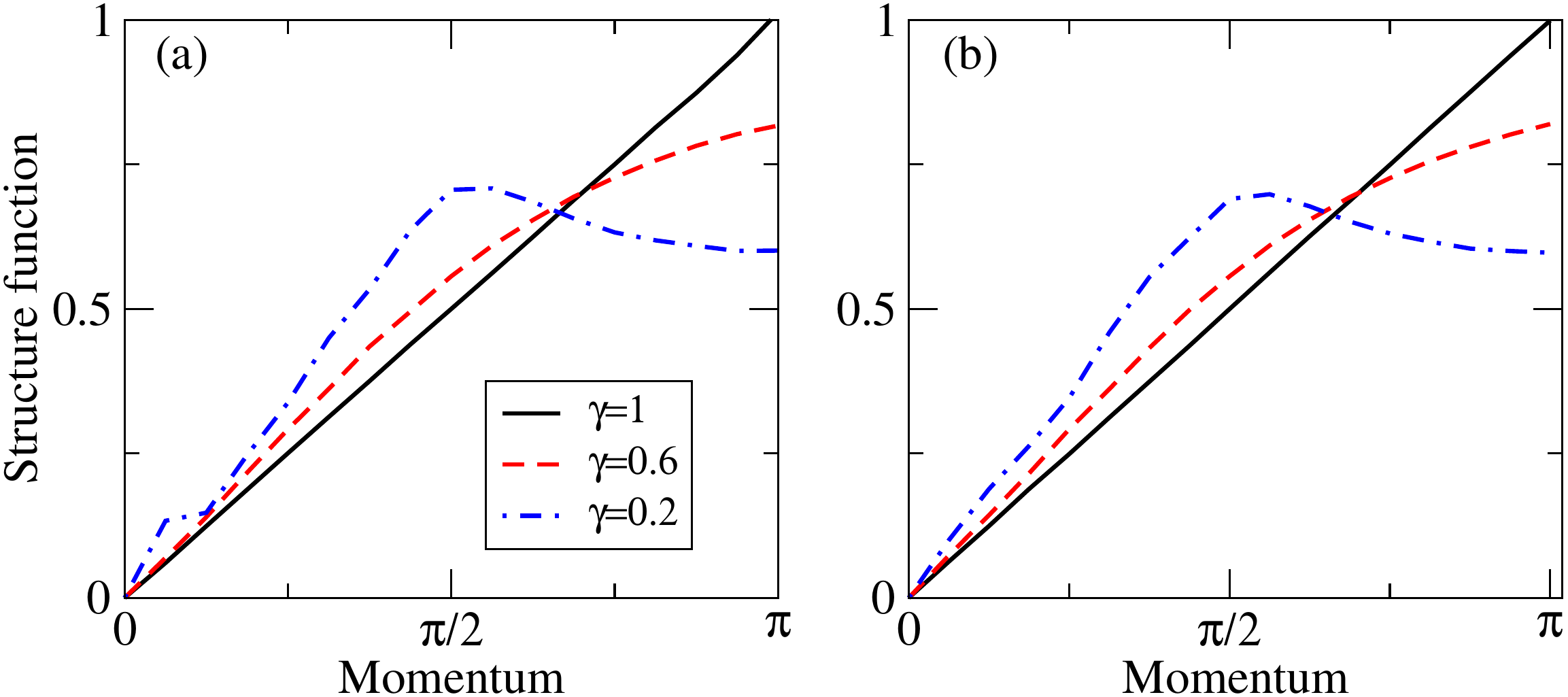}
\end{center}
\caption{(Color online) (a) Charge structure function for $U=0$.
(b) Spin structure function for $U=0$.
For the non-interacting case the charge and spin degrees of freedom
behave identically. At $\gamma=1$ they show
no structure (free fermions), but form incommensurate
ordered phases for low $\gamma$. Momentum is measured in units of the inverse lattice spacing.}
\label{mc_U0}
\end{figure}

The incommensurate order occurs generally for low values of $\gamma$ for $|U| < 4$
(region III of the phase diagram Fig.\,\ref{phdia}). Reducing $|U|$
further to $U=0$  shows the effect of $\gamma$ on a non-interacting system.
For $\gamma=1$ the system consists of free fermions, and as can be seen in
Fig.\,\ref{mc_U0} the charge and spin correlators are identical to
each other and are featureless. At $\gamma=0.6$ the dynamics of 
the system is again slightly suppressed, but at a $\gamma=0.2$
the system again manifests incommensurate charge and spin order, with
the structure functions peaking at $k = k_F$.

Let us try to understand how these results connect to the exact solution for $|U| \leq 4$. 
In this case the ground state consists
of a finite number of doublons and holons, separated by singly occupied
sites to ensure their maximal delocalization along the lattice.
Thus, in this sector a rather
special ordering, characterized by coexistence of CDW and SDW order on
different lattice sites, is possible. For a more detailed description
let us consider few particular cases.

Let us start from the $U=0$ case, where $n_d=0.25$. The ground state
configuration at $X \simeq 1$ consists of an alternating distribution
of doublons and holons, separated by single occupied sites with
alternating spins on these sites. A possible
configuration would be
$$(\,d\,\ua\,h\,\da \, d\,\ua\,h\,\da \, d\,\ua\,h
\,\da \,...)$$ showing the coexistence of period-4
charge and spin density modulations, as observed in Fig.\,\ref{mc_U0}.

At $U=-2$, where $n_d=0.3(3)$, a possible configuration would be
$$(\,d\,\ua\, h \, d\, \da \,h \, d\, \ua\, h \, d\, \da\, h\,\,
\, ...)$$ showing the  coexistence of a period-3 charge modulation
with a period-6 spin density modulation.

For other values of $U$, the number of doublons (and
singly occupied sites) will be in general incommensurate.
The structure of the coexisting charge and spin density waves 
must reflect this incommensurability, and 
will consequently be much more complicated.

For $0<U<4$, the behavior is the same as for $-4<U<0$, but with the 
spin and charge structure functions inverted. For instance, for $U=2$
Fig.\,\ref{mc_U-2}a would hold for spin and Fig.\,\ref{mc_U-2}b would 
hold for charge, indicating an incommensurate spin-charge-density wave.

\begin{figure}
\begin{center}
\includegraphics[width=0.45\textwidth,clip=true]{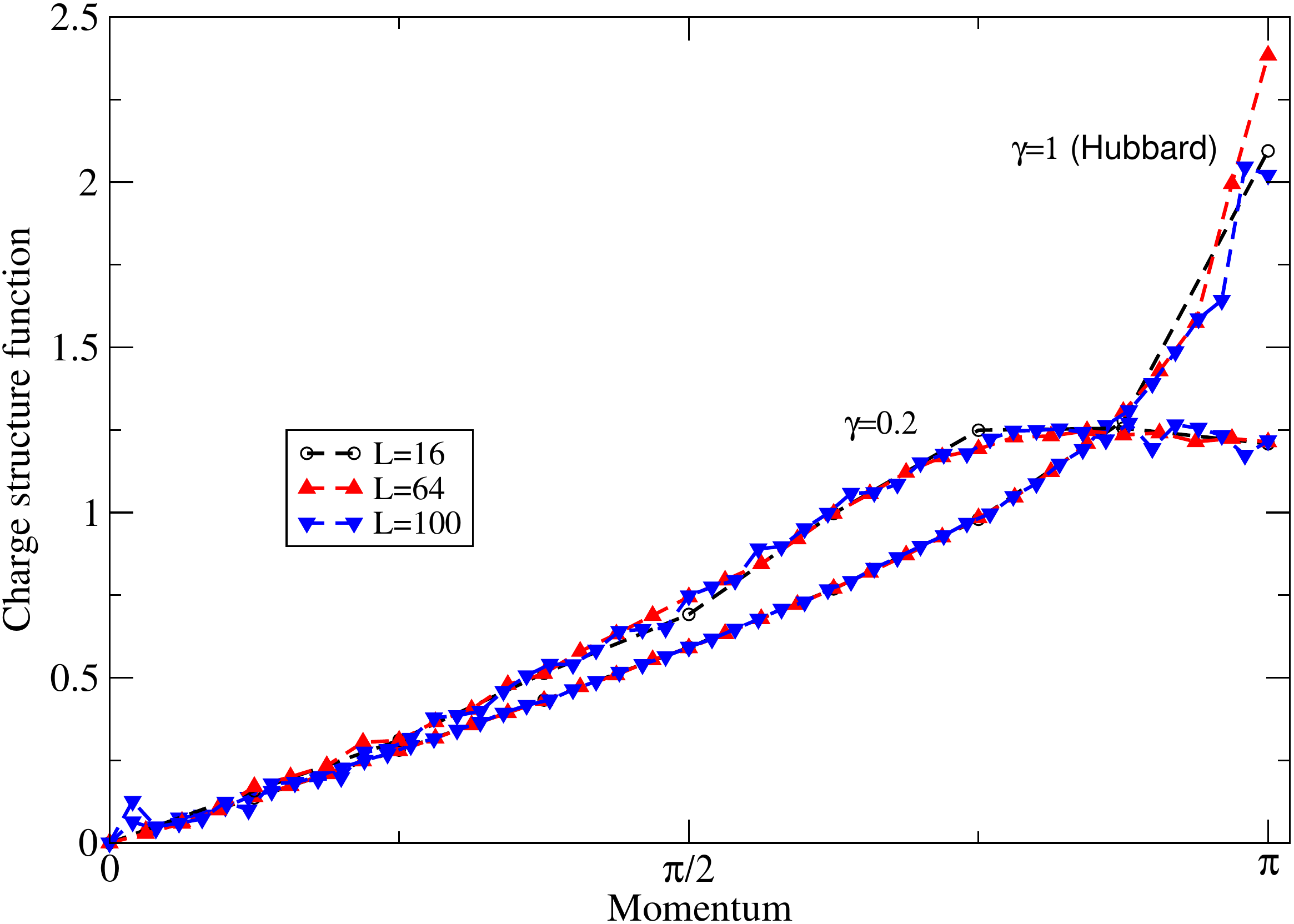}
\end{center} 
\caption{(Color online) The charge structure function for $U=-2.5$ as the lattice
size is increased. The results show little dependence on lattice
size, indicating that finite size effects are not important in our analysis. Momentum is measured in units of the inverse lattice spacing.}
\label{fss}
\end{figure}

To ensure that the behavior we have seen is not an artifact of the finite
system size, we have repeated our simulations for $U=2.5$ for
lattice sizes between 16 sites and 100 sites. We show the results in Fig.\,\ref{fss},
and it is clear that the incommensurate structure seen in the structure
functions hardly alters as the lattice size is increased. We can thus be confident that 
our standard size of $L=32$ is sufficiently large for finite size effects 
to be neglected.

\begin{figure}
\begin{center}
\includegraphics[width=0.45\textwidth,clip=true]{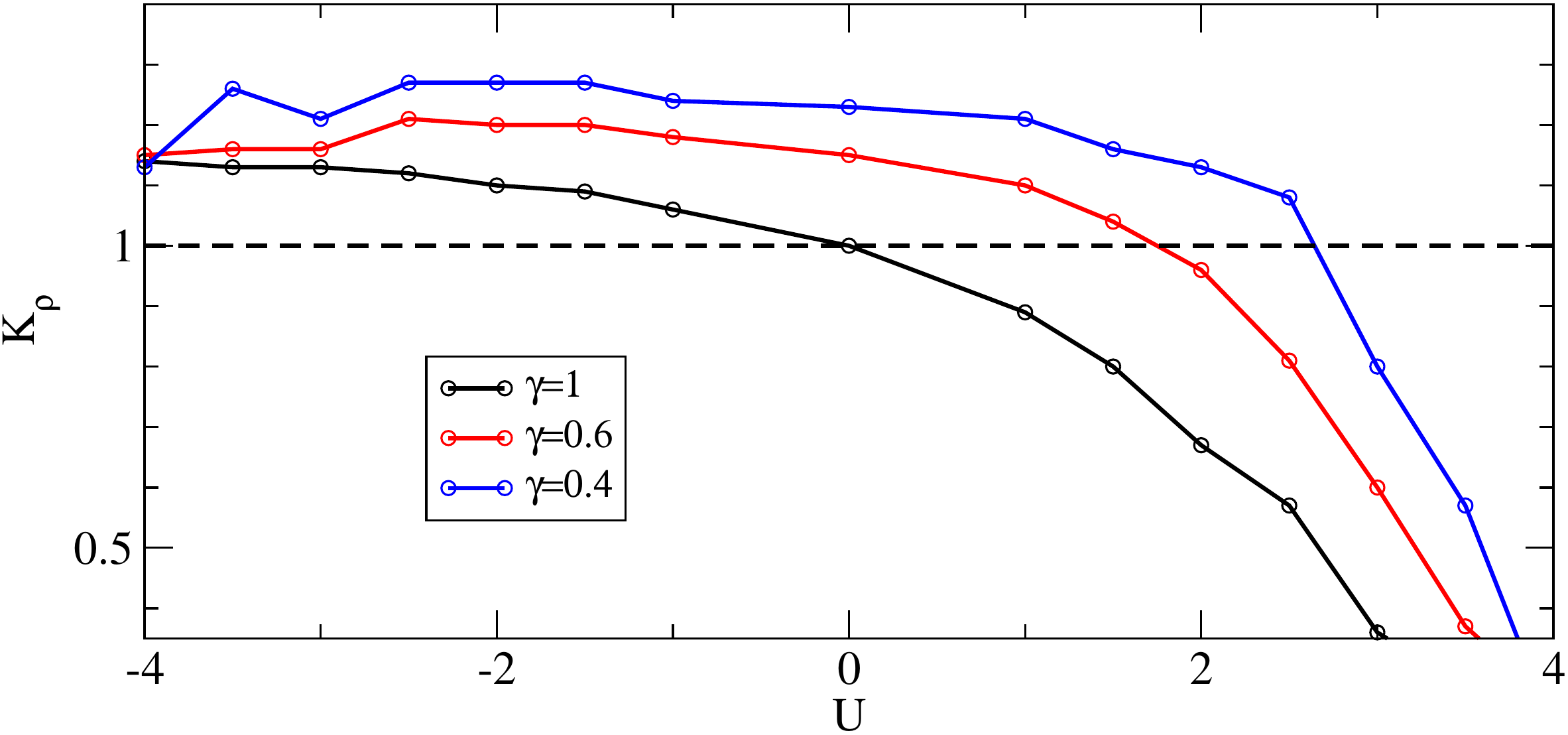}
\end{center}
\caption{(Color online) Luttinger-liquid parameter $K_\rho$. For $\gamma=1$, $K_\rho$ is
approximately equal to 1 for negative $U$, then drops as $U$ becomes positive,
signaling the opening of the charge gap. As $\gamma$ is decreased, this drop occurs at
larger values of $U$. The dashed line is a guide to the eye, to assist in
estimating where the drop occurs.}
\label{luttinger}
\end{figure}

From the simulation we have also evaluated the Luttinger parameter $K_\rho$,
using Eq.\,(\ref{lutt_par}). As shown in Fig.\,\ref{luttinger}, for $\gamma=1$
$K_\rho$ is equal to 1 for negative values of $U$ 
(the 10\% deviation is within the numerical error of our calculation), 
indicating the coexistence of superconductivity and charge wave ordering.
As $U$ becomes positive, a gap opens in the charge sector and $K_\rho$
can no longer strictly be defined for the half-filled case. This is
marked in Fig.\,\ref{luttinger} by the calculated value of
this parameter abruptly dropping, as the charge structure function
is no longer linear at small momentum. For $\gamma=0.6$ a similar
behavior is seen, except that the opening of the charge gap now occurs at
a higher value of $U \simeq 1.8$. As $\gamma$ is reduced further
this trend continues, and for $\gamma=0.2$ we find that the critical $U$ has
a value of approximately 3.5, in good agreement with the estimate given in
Ref.\,\cite{aligia}.

Before closing this section, we want to mention that
a model, similar to the one presented here, but without the
three-body term, has been been previously investigated
\cite{anfossi,montorsi}. For this so-called Hirsch model (see for
instance Ref.\,\cite{hirsch2}), the strongly-correlated regime at
half-filling exhibits an incommensurate (singlet) superconducting
phase that show many similarities with our findings. 
This phase has been captured using density-matrix renormalization group techniques (DMRG), while
bosonization and RG were unable to describe the transition \cite{dobry}. 

At present, the phase diagram of the model (\ref{Eff_Ham_1}) in 1D is only
partially known; however, as we will show in the last section, the
model can describe realistic experiments using cold atoms. This
represents a good challenge for quantum simulations, as well as for
new numerical calculations. A fascinating possibility would be an
emerging superconducting phase, with tightly bound pairs of momentum
$\pi$.


\begin{figure}[!htbp]
    \begin{center}
    \includegraphics[width=0.35\textwidth]{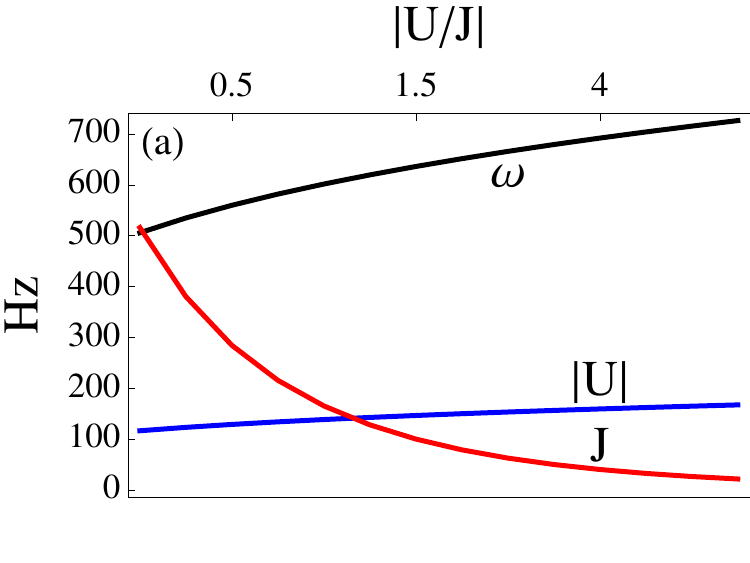}\\
    \vspace{-.5cm}
    \includegraphics[width=0.35\textwidth]{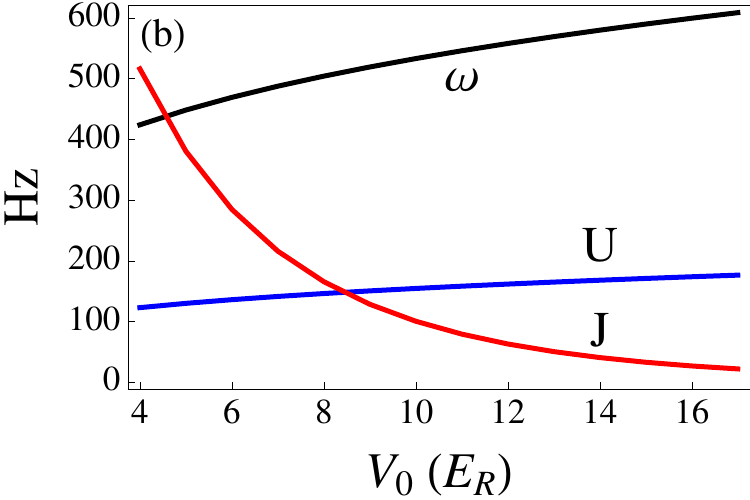}
    \end{center}
    \caption{(Color online) Comparison between frequency modulation $\om$ of the magnetic field, hopping parameter $J$ and Hubbard interaction $U$ (all measured in Hz) at the zero of the Bessel function $\mathcal{J}_0(U_1/\hbar\om)$ for ${}^{40}K$ close to a zero of a Feshbach resonance ($B_0 (G) = 224.2$, $\Delta (G) = 9.7$ and $a_{bg} / a_0 = 174$). (a) In the attractive case $a_0<0$, we have chosen $B_m (G) = 233.5$ and $B_1 (G) = 4$; (b) in the repulsive case $a_0>0$, we have chosen $B_m (G) = 234.36$ and $B_1 (G) = 4$. The optical lattices parameters are $V_x = V_y= 25\,E_R$, $V_z = V_0$ and $\lambda = 1064\rm{nm}$.}
    \label{fig:fesh}
 \end{figure}

\section{Experimental parameters.}

For the experimental realization of this model (in 1D for instance)
we consider an optical lattice $V(\textbf{r}) = V_x \sin^2(k x) +
V_y \sin^2(k y) + V_z \sin^2(k z)$ with $k=2\pi/\lambda$ generated
by a laser with wavelength $\lambda = 1064\rm{nm}$; we take the limit $V_x,\, V_y \gg V_z$ to allow
dynamics only in one dimension. We studied the Feshbach resonance for
${}^{40}K$ at $B_0 (G) = 224.2$, characterized by a width $\Delta (G) = 9.7$ and a
background scattering length $a_{bg} / a_B =
174$, where $a_B$ denotes the Bohr radius \cite{grimm}. The dependence of the scattering length $a_s$ on
the magnetic field $B$ is given near resonance by 
\be a_s = a_{bg}
\left(1-\f{\Delta}{B-B_0}\right)\,. \ee 
We choose a time dependent
magnetic field of the form $B(t) = B_m + B_1 \cos(\om t)$ and
consider $|B_1| \ll |B_m - B_0|$. Therefore, at first order in
$B_1/(B_m - B_0)$  we can write 
\bea \label{swave}
a_s &\simeq & a_{bg} \left[  1 - \f{\Delta}{B_m -B_0} \left( 1 - \f{B_1 \cos(\om t)}{B_m -B_0} \right) \right]\nn\\
&\equiv& a_0 + a_1 \cos(\om t)\,, \eea where we have defined $a_0 =
a_{bg} [1 - \Delta/(B_m - B_0)]$ and $a_1=-a_{bg} B_1\Delta  /
(B_m-B_0)^2$.

In Figs.\,\ref{fig:fesh}(a) and (b), we plot the driving frequency values
corresponding to the zero of the Bessel function
$\mathcal{J}_0(U_1/\hbar\om)=0$ both in the attractive and repulsive
case, respectively, in a particular range of parameters near the zero of the
Feshbach resonance (so that we can reach the region of interest in
the regime of strong coupling analyzed in this paper) and compare with an estimate of $U$ and
$J$ for the 1D case. We find that $\om$ is in the sub-kilohertz
regime $\om \simeq 2\pi \times 500-600 \rm{Hz}$ and we observe that
such a choice of parameters fulfills the main approximations
required from Floquet theory, \emph{i.e.} $\om\gg U,\,J$. Actually,
in typical experiments \cite{lignier2007} the kilohertz energy scale
is far below the band gap and higher band contributions do not play
a role, except for possible multiphoton processes. Moreover, one can
see that the range where Floquet theory can be applied for this
choice of parameters of the resonance allows us to explore the phase
diagram in the main region of interest, where the correlated-hopping
model should reveal interesting phenomena. Such a choice of
parameters plotted in Fig.\,\ref{fig:fesh} can be considered as
an example to show that the model described here can be realized in
experiments; one would envisage that for different values or ranges
of $U/J$, the optimal parameters will be chosen accordingly. We
finally want to mention that to calculate the parameters $U$ and
$J$ we have used the approximate formulas (given in terms of the recoil energy 
$E_R = \hbar^2(2\pi/\lambda)^2/2m$, with $m$ denoting the atomic mass)
\cite{bloch1}: $U/E_R =
(2\pi / \lambda) a_s \sqrt{8/\pi} \,V_0^{1/4}\, V_{x,y}^{1/2}$ and
$J /E_R = (4/\sqrt\pi)\, V_0^{3/4} e^{-2\sqrt{V_0}}$,
where we have introduced the potential depth $V_0=V_z$ (assuming
that the electron dynamics would be in the $z$ direction), and
frozen the motion in the $x$ and $y$ directions taking $V_x = V_y=
25\,E_R$ such that we can consider one-dimensional effective
systems.

\section{Conclusions.} 

We have discussed a scheme for cold atoms to engineer
an extension of the Hubbard model that includes nearest-neighbor correlations
affecting the hopping processes for fermions in optical lattices. After imposing a time-dependent
driving of the $s$-wave scattering length between atoms in two different hyperfine states (that we
have modeled as a pseudo-spin $1/2$ system assuming no spin imbalance), we have shown
within Floquet theory that the system can be described by an effective Hamiltonian with 
correlated-hopping interactions. The model has an additional SU(2) symmetry, with respect to the
usual spin-SU(2) symmetry of the Hubbard model, generated by the algebra of $\eta$ operators. This 
fact opens the possibility of searching for a ground state characterized by the exotic $\eta$-pairing 
superconductivity proposed by Yang in 1989 as a metastable state of the Hubbard model. 
This model, for the particular case of $d=1$ on which we focused
in this work, has two integrable points as a function of the driving parameter $X$ that tunes the 
coupling of the correlated-hopping interactions: one is the Hubbard model ($X=0$) and the other one ($X=1$) has 
been analyzed in Ref.\,\cite{aligia} by Arrachea and Aligia. The integrable point discussed by them 
manifests $\eta$-pairing in the ground state. Unfortunately, the huge degeneracy
of the ground state prevents the system from showing superconducting properties, like anomalous 
flux quantization \cite{aligiaflux}. Exploring this region of the phase diagram that extends over 
the whole filling axis (see Fig.\,\ref{phdia}, region III) can be quite challenging in general for 
experiments. Indeed, as discussed for the case of the supersymmetric
model by Essler \emph{et al.}\,\cite{essler}, it is possible to draw
the phase diagram of Fig.\,\ref{phdia} using the grand canonical ensemble. Such
representation is of fundamental importance because in typical cold atom
systems the presence of the trap can be interpreted, in the local
density approximation (LDA), as a local chemical potential such that
different shells with different quantum phases would appear radially in
the trapped gas. The consequence of this, however, is that the central
``dome'' (region III) would correspond to a single value of chemical
potential $\mu = 0$, thus rendering its observation problematic.

We have focused on the study of the half-filled model, away from the integrable
point $X=1$, using the ``world-line" algorithm to perform QMC simulations . We have explored
the parameter space in the strong coupling regime, where known analytical methods 
like bosonization and RG techniques cannot be employed. We have found that an 
incommensurate order in the charge and the spin sector
sets in for the ratio $|U/J|<4$, where $U$ and $J$ are respectively the on-site 
interaction and the bare hopping amplitude. We have observed that the two kinds of orders manifests as
a peak in the spin and charge structure functions at incommensurate (distinct) momenta. The
two orders exchange their behavior when $U\ra -U$ as expected from the symmetries of the model. 
In particular, for the case $U=0$ a peak appears exactly at $k_F=\pi/2$ in both structure functions.

A further investigation of the model would require the measurement of other types of orders, to see, 
for instance, what role is played by superconductivity when the incommensurate spin and charge
order appears. These types of correlations cannot be computed with the QMC algorithm 
used in this work since it is based on a number-conserving representation of the fermionic Hilbert space; 
one would thus need to employ other techniques to look, for instance, at the 2-body density matrix. 
Moreover, deviations from half-filling are still to be studied in the strong coupling regime and the phase 
diagram has not been established yet, except for the case $X=1/4$ \cite{nakamura}. 

In dimensions $d\geq 2$, the physics of the model is almost all to be explored; weak-coupling 
Hartree-Fock calculations in $d=2$ show that the model can exhibit $d$-wave superconductivity 
\cite{arracheadwave}. A very interesting possibility, deferred to further studies, would be the appearance of $\eta$-superconductivity 
in the ground state.

\section{Acknowledgments.} We are thankful to L. Santos for pointing
out the problem concerning the observation of phase III in trapped
systems. We also acknowledge A. Montorsi, M. Roncaglia, P.
Barmettler, M. Dalmonte, and A. Hemmerich for useful discussions,
and F. Sols for insight regarding ODLRO. This work was supported by the Netherlands Organization
for Scientific Research (NWO) and by the Spanish MICINN through
Grant No. FIS-2010-21372 (CEC). 

 \addcontentsline{toc}{section}{Bibliography}
\bibliographystyle{mprsty}
\bibliography{Biblio}

\newpage

\onecolumngrid
\appendix*

\section{Derivation of the effective model}

The Hamiltonian of the Hubbard model with a time-dependent interaction reads:
\bea
H &=& -J \sum_{\bra i,j\ket, \sigma} (c^\dag_{i\sigma} c_{j\sigma} + \rm{h.c.}) + \bar{U}(t) \sum_i n_{i \ua} n_{i\da}\nn\\
&=&  -J \sum_{\bra i,j\ket, \sigma} (c^\dag_{i\sigma} c_{j\sigma} + \rm{h.c.}) + U \sum_i n_{i \ua} n_{i\da} + U_1\cos(\om t) \sum_i n_{i \ua} n_{i\da} \nn\\
&\equiv& H_J + H_U + H_\rm{d}(t)\,.
\eea
Let us define the following Floquet basis
\be
| \left\{ n_{j\sigma} \right\}, m \ket = | \left\{ n_{j\sigma} \right\} \ket \exp\left(-i \f{U_1}{\hbar \om}\sin(\om t) \sum_j n_{j \ua} n_{j\da}  + im\om t\right)\,,
\ee
where $ | \left\{ n_{j\sigma} \right\} \ket$ stands for a Fock state, and compute the Floquet Hamiltonian matrix elements using this basis (the double brackets indicates the time average)
\be
\bra\bra \left\{ n'_{j\sigma} \right\} , m' | H-i\hbar \pa_t | \left\{ n_{j\sigma} \right\} , m \ket\ket\,.
\ee
The derivative $-i\hbar \pa_t | \left\{ n_{j\sigma} \right\} , m \ket $ cancels with $H_\rm{d} (t) | \left\{ n_{j\sigma} \right\} , m \ket$. Let us examine $ H_U$: we have to calculate the following term
\be
\f1 T \int_0^T\, dt\, e^{ i \om t (m-m') } \bra \left\{ n'_{j\sigma} \right\} | H_U | \left\{ n_{j\sigma} \right\}\ket 
\exp \left[ -i\f{U_1}{\hbar\om}\sin(\om t) \sum_j (n_{j \ua} n_{j\da} - n'_{j \ua} n'_{j\da})\right]\,.
\ee
$| \left\{ n_{j\sigma} \right\}  \ket$ are eigenstates of $H_U$, hence $n_{j\sigma} = n'_{j\sigma},\,\, \forall j$ and we find
\be
\f1 T \int_0^T\, dt\, e^{ i \om t (m-m') } \bra \left\{ n'_{j\sigma} \right\} | H_U | \left\{ n_{j\sigma} \right\}\ket = \bra \left\{ n'_{j\sigma} \right\} | H_{U} | \left\{ n_{j\sigma} \right\}\ket  \delta_{m,m'}\,.
\ee
For the hopping part, we have to calculate 
\be
\f1 T \int_0^T dt\, e^{i\om t (m-m')} \bra \left\{ n'_{j\sigma} \right\} | H_J | \left\{ n_{j\sigma} \right\}\ket \exp \left[ -i\f{U_1}{\hbar\om}\sin(\om t) \sum_j (n_{j \ua} n_{j\da} - n'_{j \ua} n'_{j\da})\right]\,.
\ee
It is crucial now to evaluate the term $\bra \left\{ n'_{j\sigma} \right\} | H_J | \left\{ n_{j\sigma} \right\}\ket $. The typical form of this quantity is
\be
\bra \left\{ n'_{j\sigma} \right\} | c^\dag_{i\sigma} c_{k\sigma}| \left\{ n_{j\sigma} \right\}\ket \,.
\ee
If $\sigma =\, \ua$ (and then define $\bar\sigma \equiv \, \da$), it implies that $n'_{i\sigma} = n_{i\sigma} + 1$, $n'_{i\bar\sigma} = n_{i\bar\sigma}$, $n'_{k\sigma} = n_{k\sigma} - 1$, $n'_{k\bar\sigma} = n_{k\bar\sigma}$ and $n'_{j\rho} = n_{j\rho}$ for $j\neq i,\, k$.
As a consequence, the density dependent part in the exponential becomes
\be
\hat s \equiv \sum_j (n_{j \ua} n_{j\da} - n'_{j \ua} n'_{j\da}) = n_{i\ua}n_{i\da} + n_{k\ua}n_{k\da} - (n_{i\ua} + 1)n_{i\da} - (n_{k\ua} - 1)n_{k\da} = -n_{i\da} + n_{k\da}\,.
\ee
An analogous result holds for $\sigma =\, \da$. We now use the integral representation of Bessel functions of first kind:
\be
\mathcal{J}_n(x) = \f{1}{2\pi} \int_{-\pi}^{\pi}\,dt\, e^{i(x \sin t - n t)}\,,
\ee
define $\tau = \om t$ and then shift $\tau \ra \tau + \pi$. The integral then becomes
\be
\f{1}{2\pi} \int_{-\pi}^\pi\, d\tau\,  e^{i (\tau + \pi) (m-m') -i\frac{U_1}{\hbar\om}  \hat s \sin(\tau + \pi)} = \f{(-1)^{m-m'}}{2\pi} \int_{-\pi}^\pi\, d\tau\,  e^{i \tau (m-m') +i\frac{U_1}{\hbar\om}  \hat s  \sin\tau }\,,
\ee
which yields
\be
(-1)^{m-m'} \mathcal{J}_{m' - m}\left( \frac{U_1}{\hbar\om} \hat s \right)\,,
\ee
that can be reabsorbed in $H_J$. In the large frequency limit $\hbar\om\gg J,U$, the off-diagonal elements of the Floquet Hamiltonian can be (perturbatively) neglected and we can thus consider only $m=m'$ and choose $m=0$ in the first Floquet Brillouin zone. Therefore, the approximate form of the Floquet Hamiltonian is 
\be
H_\rm{eff}= -J \sum_{\bra i,j\ket,\sigma} (c^\dag_{i\sigma} c_{j\sigma} + \rm{h.c.}) \mathcal{J}_0 \left[K (n_{i\bar\sigma} - n_{j\bar\sigma}) \right] +  U \sum_i n_{i \ua} n_{i\da} \,.
\ee
where we defined $K\equiv U_1/\hbar\omega$.

\end{document}